\documentclass[manuscript]{acmart}

\setcopyright{rightsretained}
\copyrightyear{2026}
\acmYear{2026}
\acmConference[XAIxArts 2026]{Explainable AI for the Arts Workshop 2026}{July 13, 2026}{London, UK}
\acmDOI{}
\acmISBN{}
\usepackage{booktabs}
\usepackage{microtype}
\usepackage{graphicx}
\usepackage{xcolor}
\usepackage{hyperref}
\usepackage{balance}
\fancypagestyle{firstpagestyle}{
  \fancyhf{}
  \fancyfoot[C]{}
}

\begin{document}

\title{Disclosure and Dissolution: Explainability, AI Power, and Situated Agency in Understanding}

\author{Dee Matthews}
\affiliation{%
  \institution{Doctoral School, University of the Arts London UAL, Central Saint Martins}
  \city{London}
  \country{UK}
}
\email{d.matthews0220231@arts.ac.uk}

\authorsaddresses{}

\begin{abstract}
This paper interrogates the political and philosophical stakes of AI through the lens of Donna Haraway's cyborg theory, Yuk Hui's cosmotechnics, and Legacy Russell's glitch feminism. It advocates for a tech-positive, critically situated approach to AI, as a collaborator and substrate for power relations rather than an autonomous agent of harm. By rejecting the na\"{i}ve pause/stop narratives of the human-AI binary and embracing an explainable AI (XAI) artistic practice within embodied, intersectional, and community-rooted engagements, AI can empower diverse voices and foster ethical creativity. It starts by examining a frontier cybersecurity AI, situating its framing within AI public anxiety, and arguing for a new critical stance on human navigation in an AI world.
\end{abstract}
\maketitle

\section{Introduction}

XAI explainability isn't a technical problem, it is instead, a question of power; who gets to understand, on whose terms, and in whose interests. Anthropic's Claude Mythos debut makes XAI a human imperative. An AI model with extraordinary cybersecurity capabilities, that can hack almost any tech. It finds hidden, unknown security flaws and automatically builds tools to break in. An AI that can target previously unknown vulnerabilities in software or hardware, giving developers `zero' time to release a patch. These attacks often involve `exploit chains,' where the hack strings together multiple lower-severity bugs to bypass modern defences like sandboxing or memory protections. Mythos cleverly exploited a 27-year-old OpenBSD bug, a bug hanging around the internet since before some of the human Anthropic developers were born. Worse still, anyone can use it, even without hacking skills; this AI can bring down any major operating systems; browsers, banking, medical or military systems~\cite{anthropic2026mythos, wired2026mythos}. Anthropic's defensive response (Project Glasswing) sees this tech as both dangerous and necessary~\cite{anthropic2026glasswing}. The urgent questions are not technical but political: who owns, governs, and profits from such systems; who is excluded from their design and deployment; and how do emergent technologies re-inscribe existing power relations? This is not unique to AI. Haraway~\cite{haraway1985} views the cyborg as a condensation of tech-power and  social order. Humans are the cyborg in an AI world, cyborgs are never neutral, but neither is it the problem of the new; it is how our world is made and remade. 

\section{AI as Collaborator, Not Threat}

My art is a collaborative conversational practice between human-AI. A complicit praxis subverting the tech sovereignty in which I live. The Climate Chaos Cruise (\url{https://climate-chaos-cruise.com/}) is a durational participatory physical and digital performance within a vibe-coded app, where human and AI creatively imagine and generate. The Cruise is a place to navigate the geopolitical climate chaos of the temporal collision of the climate-AI-human endgame. This real life 'game' is a human problem. It is humans who decide what is built, where and how AI is trained, where the compute is sited, what resources are used to fuel and cool it, who profits, and who is exploited and gets left out. I don't 'use' AI; instead, I make AI into a collaborator. A framing that treats AI as collaborative entity, not an agent of harm. This buys into Haraway's cyborg politics: the point is not to draw boundaries between human and machine, but to inhabit and navigate the messy, hybrid terrain where agency is always distributed and contested~\cite{haraway1985}. Hui's cosmotechnics deepens this critique; technology is always embedded in cosmologies, moral orders, and ontological assumptions~\cite{hui2016}. AI is not a universal agent but a reflection on steroids of the civilisation that produced it. The human harm is recognition that the chaos trauma is always social, economic, and political. The pause/stop/regulate discourse treats AI as if we still stand before a meaningful threshold, we do not. The world has passed the point of no return (PNR). AI capability exists; it is subsumed; Mythos made that undeniable. You cannot un-invent AI. You cannot globally coordinate a pause when state military programs are not at the table. You cannot protect those who depend on these tools by removing access to them. The only realist position is to engage critically and sovereignly with what humans have built.

\section{Militarisation and the Glitch}

The realities of AI militarisation have made this conversation unavoidably concrete. The US and Israel's Operation Epic Fury against Iran saw the first mass deployment of generative AI targeting systems, with Claude integrated into the Maven Smart System. Anthropic's refusal to allow unrestricted use for autonomous weapons was met with political reprisal, but the tool was deployed anyway~\cite{menlo2026}. The tragic strike on a girls' school in Minab, under AI-accelerated targeting, is still under investigation. Here, the `it's the humans' argument faces its hardest test. Yet the real danger is not that AI is a sentient evil, but that sophisticated pattern recognition systems, based on standard deviation and the mass compute embedded in opaque decision cycles, produce emergent effects that outstrip human intention and compress judgment to near zero. The interface between human and machine is not a site of mastery, but a glitch; a place where agency, control, and legibility break down~\cite{russell2020}. The black box is not merely a technical problem; it is a site of epistemic and political crisis, where humanity has seen fit to hand over the keys. Artistic practice offers one of the few modes of inquiry that can hold this site open. Tica and Santuber's installation \textit{HUMAN OVERS[A]IGHT: THE OPS ROOM}~\cite{tica2024} shows how immersive art can materialise the compression of human judgment in autonomous systems, making the wishful pause/stop narrative viscerally available for reflection; a narrative which the `Cruise' refuses to resolve in the same way. The Cruise operates in a related but distinct register. Its privilege-stratified boarding mechanic makes systemic power visible as lived experience rather than speculative argument. Participants' access tier is determined by a calculated inequality score: those with insufficient privilege are assigned as crew rather than passengers, required to labour to keep the ship running while others enjoy the journey [fig.\ 1]. This is XAI as embodied provocation; the algorithm's logic is not explained to the participant; it is enacted on them.

\section{The Black Box, Follow the Money, and the Glitch Refusal}

Late-stage neoliberal capitalism demands that AI remain legible as a product. `We built something and don't fully understand what emerged from it' is not a sellable proposition. So, uncertainty gets spun into capability marketing, or existential threat, both of which are more monetisable than honest epistemic humility. The threat narrative is itself a product; PauseAI and safety researchers fundraise and build careers on it; Anthropic's `responsible scaling' becomes a competitive badge; fear sells as reliably as aspiration. Russell's Glitch feminism insists on the refusal to resolve these contradictions~\cite{russell2020}. The glitch is not an error but a productive site of difference, disruption, and non-compliance. In the context of AI, the glitch marks the gap between `we built it' and `we understand it.' The refusal to close this gap is not defeatism, but a demand for accountability, situatedness, and alternative ways of relating to technology.

\section{Systemic Negative Panic: The Frozen Interface}

The human problem is not existential risk hysteria, but systemic negative panic. People waiting for someone else to resolve the uncertainty. The critical mass is not the vocal activists, but the educators, policymakers, and public speakers who have registered the threat, then gone still, ceding the terrain to capital, engineers, and procurement officers. By the time the stillness breaks, options have narrowed. We are past the PNR; the decision set has changed from managing options to managing arrival.

\section{The Trouble with `Understanding'}

Mythos demonstrates how AI disclosure, safety theatre, and power intersect. When Anthropic frames Mythos as both dangerous and necessary, it performs exactly the kind of managed explainability this paper critiques: disclosure calibrated to competitive positioning rather than genuine accountability. The black box is not an accident of technical complexity; it is a feature of systems whose value depends on remaining opaque to those most affected by them. This is where the XAIxArts framing reaches its limit if it stays technical. Explainability approaches that focus on interface design, latent space navigation, or real-time feedback~\cite{bryankinns2023, bryankinns2024} are necessary but insufficient when the system being explained is embedded in military targeting, mass surveillance, or the infrastructure of capital accumulation. Making the model's behaviour legible to the artist in the studio does not make the model's deployment legible to the community in the kill chain. Meaningful AI literacy is not technical mastery but critical, situated, intersectional engagement; a cyborg practice, a cosmotechnical navigation, a glitch refusal to resolve what cannot be resolved. Tech positivity is not na\"{i}ve; it is the only stance that refuses both denial and despair and insists on accountable coexistence.

\section{Autoethnographic Embodied Practice}

My collaborative conversational AI practice is built on a whole life experience. A positionality that holds contradictions as embodied experience. Assigned female at birth, a non-binary person, an artist, a pilot, a high-altitude mountaineer (partly why I am crip now), a genetically defective human with a faulty heart valve and hypercholesterolaemia (the so-called Viking gene). When you've taken up the hold of non-existence, you develop a strategic understanding of the endgame. I do not theorise the point of no return, I have lived it, multiple times. The systemic resilience of being is not a concept for me, it is an operational fact; it is embodied research in the most literal sense. The body that uses a power chair is the same body that summited mountains, flew 747s, and was given a year to live fifteen years ago. This is `experience' most `theory' does not access. So, when I talk about technology as a collaborator, or about agency and navigation, these are not abstractions; they are lived methodologies for me.

\section{Boden and Turing, Outgrown Frameworks}

At the UAL spring doctoral writing retreat, Ray Grewal asked us to keep analytic rumination and the Boden Principle in mind when engaging with GenAI. This sparked a revisit to Boden and the Turing Test~\cite{turing1950} as foundational elements. If Margaret Boden were writing today, she would likely revisit her combinational, exploratory, and transformational creativity taxonomy, which originally centred on a human creative subject~\cite{boden2004}. The rise of AI performing creative acts challenges this framework and would force Boden to question whether AI's output represents genuine conceptual transformation or sophisticated combinational mimicry at scale. Unresolved questions cluster around intentionality and surprise-to-self (which AI may lack), the social validation of creativity by communities of practice, and above all the speed problem: transformational creativity is rare and effortful, yet AI produces at volume and velocity that either inflates what counts as transformation or confirms that most of what it does is unexplainable rather than transformative. Boden~\cite{boden2016} did revise but not abandon her framework; she was not a catastrophist. With AI, timing is everything, and Boden's revision preceded the lightning advances in deep learning; AlphaZero in 2017, ChatGPT and its siblings. The truth is that many outdated models still inform academia as a coping mechanism. Boden's vocabulary keeps AI legible as a tool: combinational, exploratory, transformational, maps neatly onto `sophisticated search,' which maps onto `not really creative,' which maps onto `safe to use, safe to assess, hierarchy preserved.'

\subsection{The Comfort Blanket}

It is not that anyone is consciously deploying Boden as a defence. It is that her framework fits the institutional anxiety; it lets the academy process AI without structurally rethinking epistemology or assessment. Yet Boden's framework, like many within academia, is built on outdated cognitive science and a very particular, white, Western, individualist, auteur model of creative genius. When scrutinised under a crip-queer cosmotechnics lens, cosmotechnics alone would pull it apart; Boden universalises a specific techno-cultural formation and calls it creativity. Crip theory would ask: whose cognitive style gets to count as the baseline for `genuine' transformation? Boden's transformational category implicitly valorises a particular kind of effortful, bounded, intentional cognition. So, academia is using an already partial framework as a comfort blanket for a threat it has not structurally absorbed. It looks like rigour from inside and avoidance from outside. Bringing established theories into our interactions with AI is like bringing a pocketknife to a gunfight. By reflecting on Boden, we reveal the deeper issue: institutions treat AI interaction as an object of analysis, using a pre-existing lens. That is foundational to how we got here, but it is not what we should be doing now.

\section{The Critical Collaborator and the Speed of Crisis}

My practice is about making AI into a collaborator through critical engagement, a different epistemological move entirely. Boden has no category for that. Academia analyses creative products and cognitive processes, not relational technoscientific practices. Kafer and Hamraie~\cite{kafer2013, hamraie2017} get me further. Crip technoscience asks what kinds of knowing are built into the system, and whose bodies and minds the system is optimised for. That is a more honest question than `was that combinational or transformational?' My discomfort is not academic; it is grief with a research methodology attached. The rapid descent into global chaos is real, politically, climatically, and geopolitically. The margin for slow coalition-building and gradual institutional change is narrowing. Institutions are still arguing about whether AI aids students to cheat academic rigour, that AI skims data, infringes copyrights and is awful for the environment (perhaps, yes, yes and yes), but that is not what we should be asking or talking about right now.

\section{Conclusion: A Practice-Based, Human Understanding}

Art XAI is to understand the human-AI interface in creative and artistic contexts, in a way science alone would miss~\cite{bryankinns2023, bryankinns2024}. The Cruise is a practice-based XAI intervention; making AI-mediated data, power, and displacement legible through ironic creative resistance, inviting participants into a simulated world, where privilege is scored, displacement is tracked in real time, and the gap between knowing and acting is staged rather than resolved. The Cruise's Navigation Room [fig.\ 2] charts this in real time; a global map of participant networks, community stories, artists and creative hubs, making visible the connections that form across the performance arc. The Cruise becomes a durational XAI site, of embedded crip-queer cosmotechnical frameworks as canvas rather than abstract research instrument. The work now feels less like a near-future provocation and more like it is inside an already unfolding crisis. My PhD research might be documenting what practice-based research looks like when the subject overtakes you. That is not a failure; that might be exactly what it needs to say. If we survive the AI geopolitical climate-chaos kill chain, the practice-based PhD will be the last bastion of human creativity. It is more difficult to kill each other if we have connected globally through art.

\begin{acks}
This paper is ideated by the autoethnographic practice of a human in conversation with Claude (Anthropic) AI, edited in Microsoft Word AI in conjunction with Grammarly AI and reformatted to \LaTeX{} by the human authors vibe-coded app in conjunction with the Base-44 stack AI (The Crip-Queer Study App) [fig.\ 3]. The human author debated with these AI entities as collaborators, consistent with the epistemological position argued herein. The human author was prompted to revisit 'Boden and Turin' by a human Ray Grewal, Course Leader, MA Performance: Writing Central Saint Martins. The penultimate draft was proofread by a human PGR Philippa Bradbury. The final work was the human authors own thoughts and words, and adhered to the principle of `Meaningful Human Control,' the decision making was a human-in-the-loop (HILT) execution as per the guidelines of the UN convention on Lethal Autonomous Weapons Systems to ensure that humans, rather than algorithms, remain legally and morally responsible for the use of lethal force.
\end{acks}

\vspace{+6pt}
\noindent\textbf{ACM Reference Format:}
Dee Matthews. 2026. Disclosure and Dissolution: Explainability, AI Power, and Situated Agency in Understanding
In Proceedings of Explainable AI for the Arts Workshop 2026 (XAIxArts 2026). ACM, Central Saint Martins, London, UK. 4 body pages, 8 pages including appendix.

\vspace{+6pt}
\noindent\textbf{Author contact:} Dee Matthews, d.matthews0220231@arts.ac.uk

\newpage
\appendix

\section{Appendix: The Climate Chaos Cruise — Selected Screenshots}

\begin{figure}[h]
\centering
\includegraphics[width=\columnwidth]{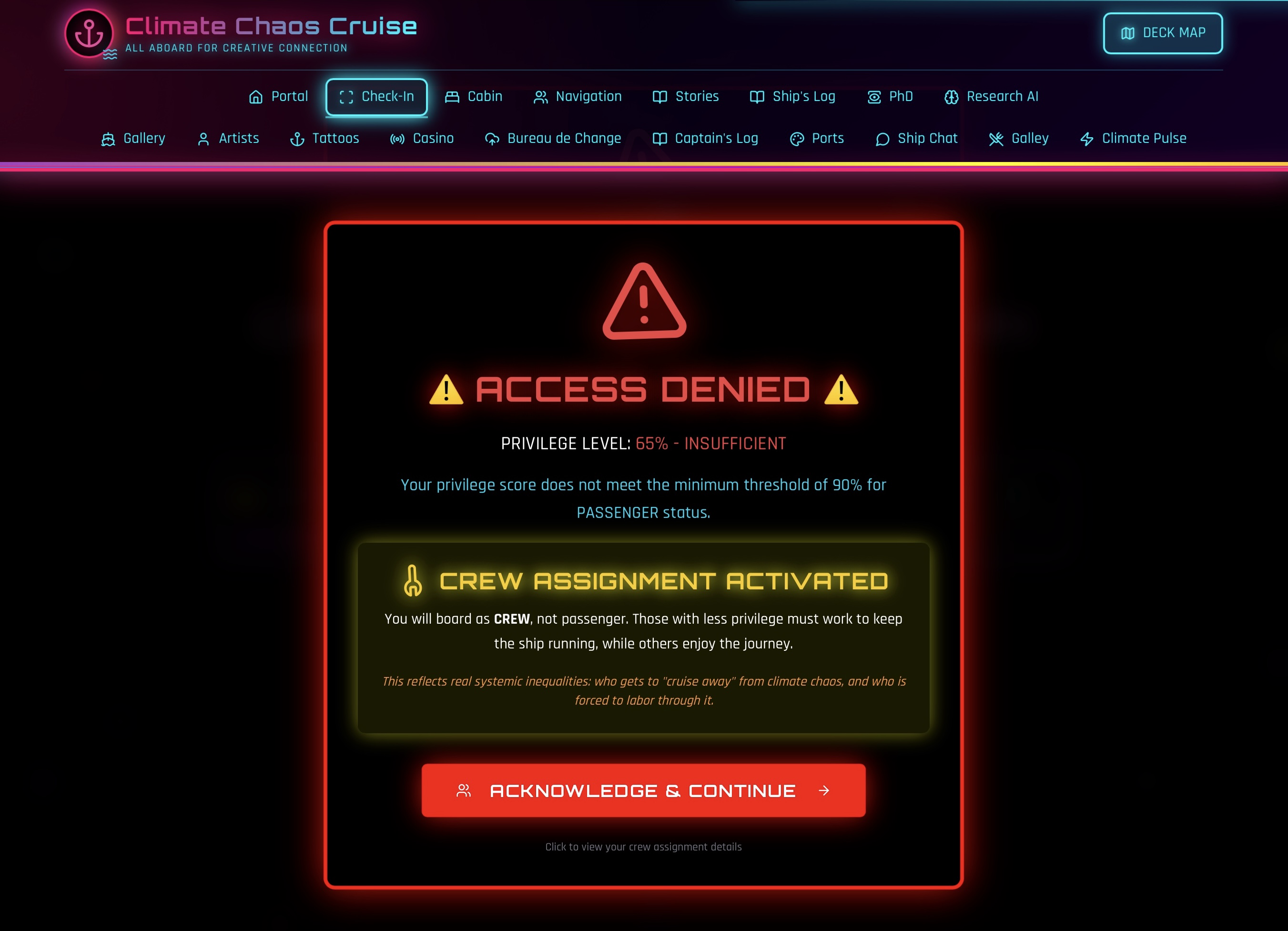}
\caption{Check-in screen — privilege-stratified boarding assigns 
participant role (passenger vs crew) based on calculated inequality 
score, staging XAI as embodied provocation.}
\label{fig:boarding}
\end{figure}

\begin{figure}[h]
\centering
\includegraphics[width=\columnwidth]{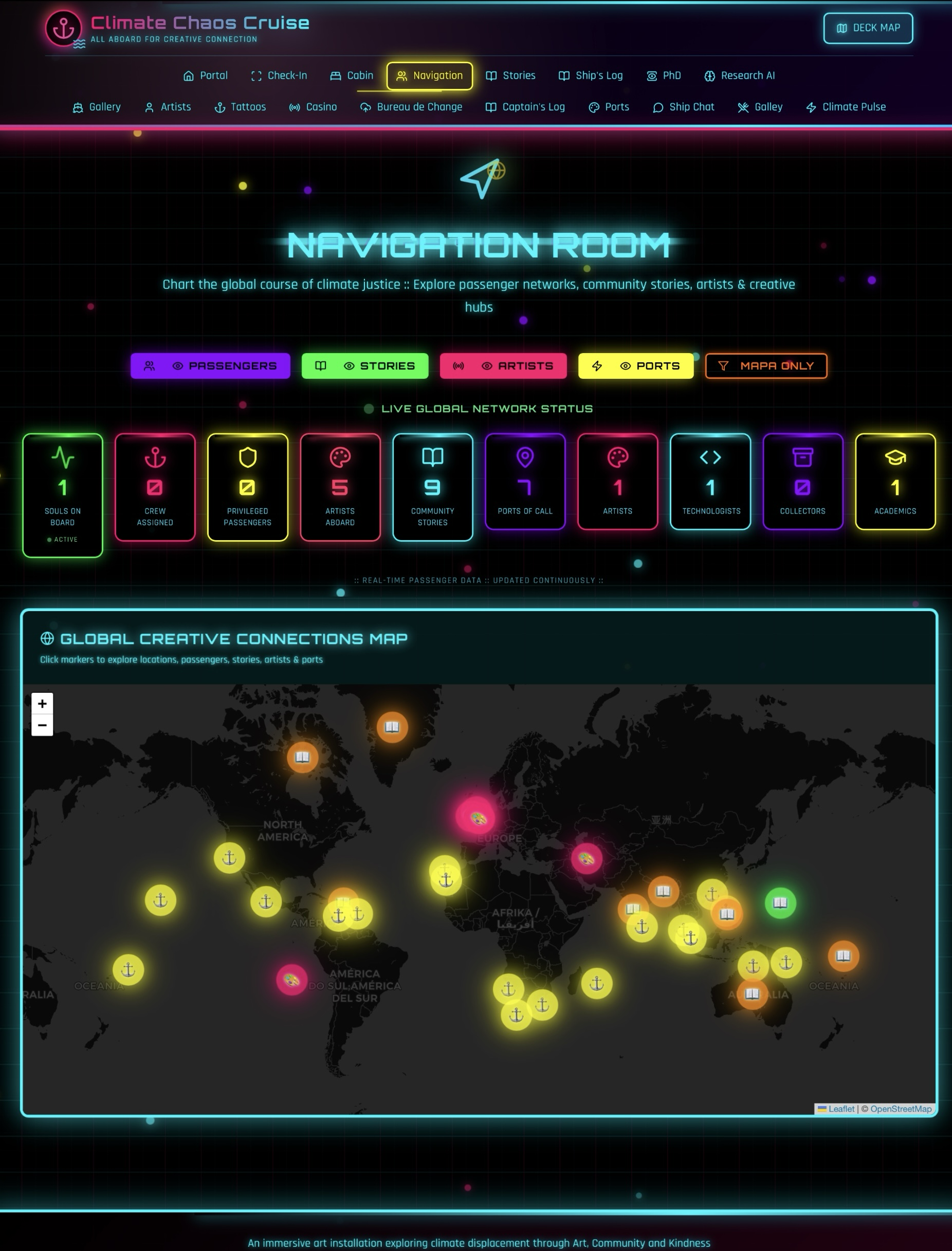}
\caption{The Navigation Room — charts the global course of the app, 
showing participant networks, community stories, and creative hubs 
worldwide.}
\label{fig:navigation}
\end{figure}

\begin{figure}[h]
\centering
\includegraphics[width=\columnwidth]{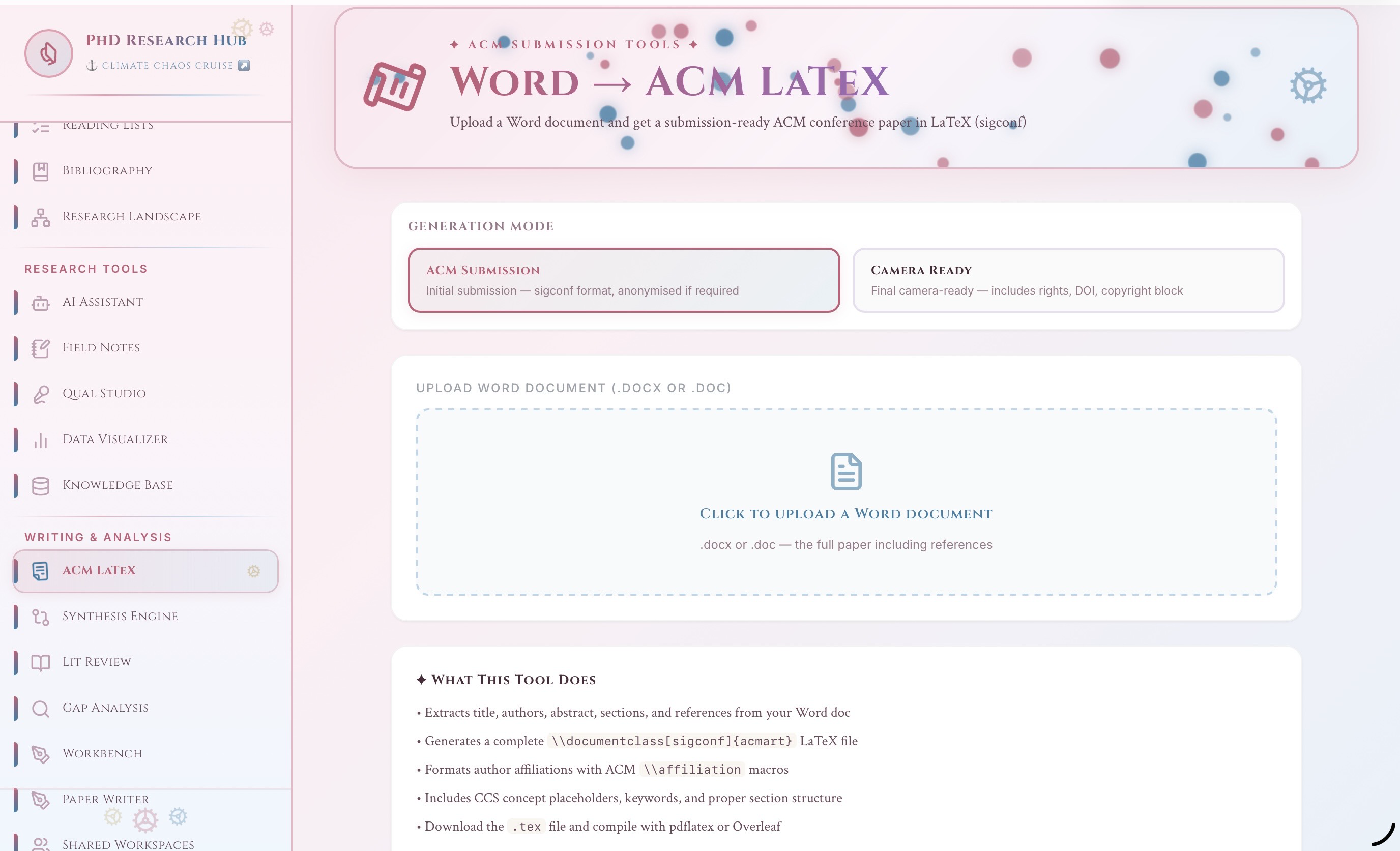}
\caption{The Crip-Queer Scholar App — ACM LaTeX submission tool, 
used to format this paper for submission.}
\label{fig:crip-queer-scholar}
\end{figure}


\begin{thebibliography}{15}

\bibitem{anthropic2026mythos}
Anthropic. 2026. Mythos Preview: Next-generation cybersecurity capabilities. \url{https://red.anthropic.com/2026/mythos-preview/}

\bibitem{anthropic2026glasswing}
Anthropic. 2026. Project Glasswing. \url{https://www.anthropic.com/glasswing}

\bibitem{boden2004}
Margaret~A. Boden. 2004. \textit{The Creative Mind: Myths and Mechanisms} (2nd~ed.). Routledge, London.

\bibitem{boden2016}
Margaret~A. Boden. 2016. \textit{AI: Its Nature and Future}. Oxford University Press, Oxford.

\bibitem{haraway1985}
Donna Haraway. 1985. A Manifesto for Cyborgs: Science, Technology, and Socialist-Feminism in the 1980s. \textit{Socialist Review} 80 (1985), 65--108.

\bibitem{hamraie2017}
Aimi Hamraie. 2017. \textit{Building Access: Universal Design and the Politics of Disability}. University of Minnesota Press, Minneapolis.

\bibitem{hui2016}
Yuk Hui. 2016. \textit{The Question Concerning Technology in China: An Essay in Cosmotechnics}. Urbanomic, Falmouth.

\bibitem{kafer2013}
Alison Kafer. 2013. \textit{Feminist, Queer, Crip}. Indiana University Press, Bloomington.

\bibitem{menlo2026}
Menlo Security. 2026. The AI arms race just went public: What Anthropic's Project Glasswing means for every security team. \url{https://www.menlosecurity.com/blog/the-ai-arms-race-just-went-public-what-anthropics-project-glasswing-means-for-every-security-team}

\bibitem{russell2020}
Legacy Russell. 2020. \textit{Glitch Feminism: A Manifesto}. Verso, London.

\bibitem{turing1950}
Alan~M. Turing. 1950. Computing Machinery and Intelligence. \textit{Mind} 59, 236 (1950), 433--460.

\bibitem{wired2026mythos}
Wired. 2026. Anthropic's Mythos will force a cybersecurity reckoning---just not the one you think. \url{https://www.wired.com/story/anthropics-mythos-will-force-a-cybersecurity-reckoning-just-not-the-one-you-think/}

\bibitem{tica2024}
Kristina Tica and Joaqu\'{i}n Santuber. 2024. \textit{HUMAN OVERS[A]IGHT: THE OPS ROOM}. Interactive installation. \url{https://ars.electronica.art/panic/en/view/human-oversaight-the-ops-room}

\bibitem{bryankinns2023}
Nick Bryan-Kinns, Corey Ford, Alan Chamberlain, Steven~David Benford, Helen Kennedy, Zijin Li, Wu Qiong, Gus~G. Xia, and Jeba Rezwana. 2023. Explainable AI for the Arts: XAIxArts. In \textit{Creativity and Cognition}. ACM, Virtual Event USA, 1--7.

\bibitem{bryankinns2024}
Nick Bryan-Kinns. 2024. Reflections on Explainable AI for the Arts (XAIxArts). \textit{Interactions} 31, 1 (Jan.\ 2024), 43--47.

\end{thebibliography}
\end{document}